\documentclass[10pt,floatfix]{revtex4-1}

\usepackage{bm}
\usepackage{color}
\usepackage{subfig}

\usepackage{amssymb,amssymb,amsfonts,amsmath,empheq,lastpage,bm,textcomp}
\usepackage[margin=1.in]{geometry}

\usepackage{caption}
\usepackage{epstopdf}
\usepackage{natbib}
\usepackage{setspace}
\usepackage{dcolumn,helvet,times}
\usepackage{graphicx}

\def\d{\mathrm{d}}

\newenvironment{Figure}
  {\par\medskip\noindent\minipage{\linewidth}}
  {\endminipage\par\medskip}

\begin{document}

\title{Motor regulation results in distal forces that bend partially disintegrated \textit{Chlamydomonas} axonemes into circular arcs.}

\author{V. Mukundan$^{*+}$, P. Sartori$^{\dagger+}$, V. F. Geyer$^{*\&}$, F. J\"{u}licher$^{\dagger\S}$, and J. Howard$^{*\&\S}$}

\affiliation{*Max Planck Institute of Cell Biology and Genetics, Dresden, Germany. $^\dagger$Max Planck Institute for the Physics of Complex Systems, Dresden, Germany. $^{\&}$Department of Molecular Biophysics \&  Biochemistry, Yale University, New Haven, CT,  USA. $^+$Both these authors contributed equally to this work.  $^{\S}$To whom correspondence should be addressed. E-mail: julicher@pks.mpg.de and jonathon.howard@yale.edu}

\begin{abstract}
The bending of cilia and flagella is driven by forces generated by dynein motor proteins. These forces slide adjacent microtubule doublets within the axoneme, the motile cytoskeletal structure. To create regular, oscillatory beating patterns, the activities of the axonemal dyneins must be coordinated both spatially and temporally. It is thought that coordination is mediated by stresses or strains, which build up within the moving axoneme, and somehow regulate dynein activity. While experimenting with axonemes subjected to mild proteolysis, we observed pairs of doublets associate with each other and form bends with almost constant curvature. By modeling the statics of a pair of filaments, we show that the activity of the motors concentrates at the distal tips of the doublets. Furthermore, we show that this distribution of motor activity accords with models in which curvature, or curvature-induced normal forces, regulates the activity of the motors. These observations, together with our theoretical analysis, provide evidence that dynein activity can be regulated by curvature or normal forces, which may, therefore, play a role in coordinating the beating of cilia and flagella.
\end{abstract}

\maketitle

\section*{INTRODUCTION}
Cilia and flagella are long, thin motile organelles containing an axoneme. The axoneme, in turn, contains nine microtubule doublets, a central pair of microtubules, motor proteins in the axonemal dynein family, and a large number of additional structural and regulatory proteins \citep{pazour_proteomic_2005}. Cilia and flagella undergo regular oscillatory beating patterns that propel cells through fluids and propel fluids across the surfaces of cells. The flagellar beat is powered by the dyneins, which generate sliding forces between adjacent doublets.
%t
 This sliding is converted to bending by constraints at the base of the axoneme (the basal body located in the cell body) and/or along the length of the axoneme (nexin links) \citep{alberts_molecular_2002, brokaw_direct_1989}. 
When these  constraints are removed by proteolysis, sliding between doublets leads to  telescoping of the axoneme up to 9 times its initial length \citep{summers_adenosine_1971}.

While the constrained-sliding mechanism of bend formation is understood, it is not known how the activities of the dyneins are coordinated in space and time to produce the beating pattern. It is thought that the 
beat is the result of feedback. The axonemal dyneins generate forces, deforming the axoneme; the deformations, in turn, regulate the dyneins.
Because of the geometry of the axoneme, deformation leads to stresses and strains that have components in various directions (e.g. axial and radial). Which component regulates the dyneins is not known.

Three different, but not mutually exclusive, models for dynein regulation have been suggested in the literature. 
According to the sliding-control model, dyneins are regulated by tangential forces acting parallel to the long axis of the microtubule doublet \citep{brokaw_molecular_1975, julicher_spontaneous_1997, camalet_generic_2000,riedelkruse_how_2007}. 
According to the curvature-control model, dyneins are regulated by doublet curvature \citep{brokaw_bend_1971, brokaw_thinking_2009}. 
According to the transverse-force model, dyneins are regulated by forces normal to the doublet surface \citep{lindemann_model_1994}. Which of these mechanisms applies to axonemes is not known.  

In this work, we partially disintegrated axonemes using protease treatment in order to study the interaction of pairs of doublets. In the course of these experiments, we made the surprising observation that at low ATP concentration, the doublets can bend into circular arcs with almost constant curvature. By developing a theoretical model of the statics of a pair of doublets, we deduce that the sliding forces must be concentrated at the distal tips of the doublets. Furthermore, we show that the curvature and transverse-force models, but not the sliding model, can readily account for this distribution of motor activity.

\section*{MATERIALS AND METHODS}
\section*{Statics of bent filament pairs}

We model the pair of adjacent microtubule doublets studied in our experiments as two inextensible filaments crosslinked by motor proteins and elastic elements (Figure \ref{fig:schematic}). The model, which is similar to previous models \citep{camalet_generic_2000, riedelkruse_how_2007}, is two dimensional; in this case, stress, which is force density, has units of force per unit length rather than force per unit area in the three-dimensional case. We characterize  the  shape of the filament pair by the tangent angle $\psi(s)$,  a function of the arc-length $s$ measured from the base where $s=0$, along the centerline between the filaments. We use a reference frame for which  $\psi(0)=0$ when the system reaches its final, quasi-static state of association (e.g. $5.7\,{\rm s}$ in Figure \ref{fig:bendingmont}A).  The filaments are separated by a distance  $a(s)$, { the sum of the minimum distances between the centerline and each of the two filaments,} which may depend on arc-length. For a given shape of the filament pair, the local shear displacement $\Delta(s)$ of one filament with respect to the other (Figure \ref{fig:schematic}) is
\begin{equation}
\label{slidingd}
\Delta (s) = \Delta_{\rm{0}} + \int_0^s\ a(s') \dot{\psi}(s')\, \d s'\quad ,
\end{equation}
where $\Delta_{\rm{0}}$ is the sliding displacement at the base { (see discussion after Eq. \ref{basvar})}. The dot denotes differentiation with respect to arc length, so that $\dot{\psi}(s)$ is curvature, also denoted $C(s)$. The sign convention is defined by Figure 1 and summarized in Appendix B.

\begin{Figure}
\begin{center}
\includegraphics [width=8.6cm] {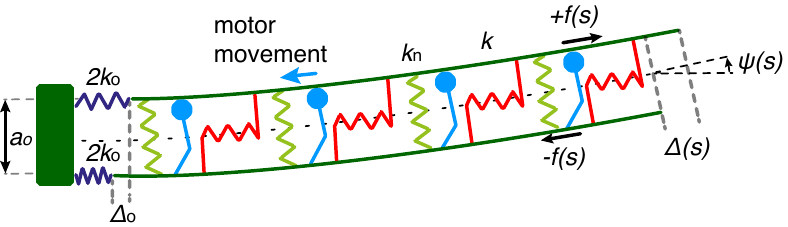}
\captionof{figure}{Scheme of two microtubule doublets bent by motors, as seen in the bending plane. The two filaments (dark green) are constrained at the base with a spacing $a_0$. The dyneins (light blue) step towards the base of the doublets. This produces a force density on the top filament $+f(s)$ putting it under tension, which slides it towards the distal end; and a compressive force density $f(s)$ on the bottom filament. The local sliding displacement is given by $\Delta(s)$, and the sliding at the base is $\Delta_{\rm 0}$. The red springs denote the shear stiffness $k$ of cross-linking elements, the green springs denote the normal stiffness $k_{\rm n }$ and the two blue springs denote the basal stiffness $2k_{\rm{0}}$ of each doublet. The tangent angle at the arc length position $s$ is denoted by $\psi (s)$. The sign convention is such that the dyneins generate a positive force density ($f(s)>0$) which in turn produces a positive angle and a positive curvature as shown in the figure.}
\label{fig:schematic}
\end{center}
\end{Figure}

To deduce the static mechanical properties of the filament pair shown in Figure \ref{fig:schematic}, we take a variational approach. 
We calculate the shape and separation that minimize the mechanical energy of the filament pair, including the work done by the motors.
 The mechanical energy of the filament-motor system is:
\begin{align} \label{FreeEnergy}
U &= \int_0^L \ \bigg\{ \dfrac {\kappa} {2}\dot{\psi}(s)^2 +\dfrac{k}{2}\Delta(s)^2
 +  \dfrac{k_{\rm{n}}}{2}(a(s)-a_{\rm{0}})^2\nonumber\\
 &- f_{\rm{m}}(s)\Delta(s) \bigg\}\, \d s + \frac{k_{\rm 0}}{2}\Delta_{\rm 0}^2
\end{align}
$\kappa$ denotes the bending rigidity of the filament pair. {$k$ and $k_{\rm n }$ denote the stiffnesses per unit length of tangential and normal springs that oppose, respectively, sliding and separation between doublets. Candidates for these springs include the nexin links \citep{Lindemann2003} and the dynein motors themselves \citep{Okuno1980}.}  $a_{\rm 0}$ denotes the unstressed filament separation and $f_{\rm m}(s)$ denotes the force  per unit length generated by the motors. Finally, $k_{\rm 0}$ denotes the effective tangential stiffness at the base.

The variations of $U$ with respect to  $\psi(s)$,  $a(s)$ and  $\Delta_{\rm 0}$ are calculated in Appendix A. For the variation with respect to the angle we obtain:
\begin{equation}\label{dgdr}
\frac {\delta U}{\delta \psi} = -\kappa \ddot{\psi}(s) - a(s)f(s) + \dot{a}(s)F(s)\quad ,
\end{equation}
where we have introduced the shear force per unit length  $f(s)=f_{\rm m}(s)-k\Delta(s)$, and  the total shear force  $F(s)=\int_s^Lf(s')\d s'$.  The variation with respect to filament separation is:
\begin{equation}\label{dgda}
\frac {\delta U}{\delta a}=f_{\rm n}(s) - \dot{\psi}(s)F(s)\quad ,
\end{equation}
where $f_{\rm{n}}(s)=k_{\rm{n}} (a(s)-a_{\rm{0}})$ is the normal force per unit length. {In this model, the normal force is the important quantity: change in spacing ($a(s)-a_{\rm 0}$) and normal stiffness  $k_{\rm n}$  can be chosen arbitrarily, provided that their product equals the normal force. For example, if the normal stiffness is very high, then the change in spacing is very small, and in this case we can think of the spacing being constant along the whole length.}

 Finally, the variation with respect to the basal sliding displacement is:
\begin{equation}
\label{basvar}
\frac {\partial U}{\partial \Delta_{\rm 0}}=F_{\rm{0}} - \int_0^L f(s) \d s\quad ,
\end{equation}
where the force acting at the base is $F_{\rm 0} = k_{\rm 0} \Delta_{\rm 0}$. {The basal force F$_0$ is the important quantity. Using similar reasoning to that used with respect to the normal force, $\Delta _{\rm 0}$ and $k_{\rm 0}$ can be chosen arbitrarily, provided that their product equals the basal force. For example, if the basal stiffness is very high, then $\Delta _{\rm 0}$ is very small, and in this case there is effectively no basal sliding.}

At mechanical equilibrium in a static (non-moving) system, the forces balance and the variations in the energy vanish: $\delta U/\delta\psi=0$, $\delta U/\delta a=0$ and $\partial U/\partial \Delta_{\rm 0}=0$. For $\partial U/\partial\Delta_{\rm 0}=0$ it follows from Eq. \ref{basvar} that the total shear force is balanced at the base, $F_{\rm 0}= \int_0^Lf(s)\d s$.
  
For $\delta U/\delta a=0$ we obtain from Eq. \ref{dgda}  the balance of normal forces \citep{lindemann_model_1994}:
\begin{equation}
\label{t-force}
f_{\rm n}(s)=\dot{\psi}(s)F(s)\quad .
\end{equation}
This  shows that the normal force per unit length $f_{\rm n}$ arises from the interplay of curvature and shear. For $\delta U /\delta \psi=0$, Eq. \ref{dgdr} provides an equation that describes force-balanced filament shapes.

In the following, we consider the limiting case where the normal springs are stiff ($k_{\rm n}$ large) and consequently $a(s)= a_0$. In this limit, the normal force $f_{\rm n}(s)$ plays the role of a Lagrange multiplier to impose the constraint of fixed inter-doublet spacing. Integrating Eq. \ref{dgdr} for $\delta U /\delta \psi=0$ and with boundary conditions corresponding to no external torques ($C(L)=0$, see Appendix A), we obtain the moment balance $\kappa C(s)=a_0F(s)$. Substituting this into Eq.\,\ref{t-force} we obtain the normal force
\begin{align}
\label{n-force}
f_{\rm{n}}(s)=\kappa C(s)^2/a_0\quad .
\end{align}
Because $f_{\rm n}(s)$ is always positive, curvature always  tends to separate the two filaments.

\section*{Experimental procedure}

\textit{Chlamydomonas reinhardtii} cells (CC-125 wild type mt+ 137c, 

 R.P. Levine via N.W. Gillham, 1968) were grown in 2 liter TAP+P buffer for 3 days under conditions of constant illumination (2x75 W, fluorescent bulb) and air bubbling at $24\,^{\circ}\mathrm{C}$ to a final density of $10^6\,{\rm cells/ml }$ \citep{alper2012reconstitution}. Flagella were isolated using  dibucaine  \citep{witman_chlamydomonas_1978}. Purified flagella were then resuspended in HMDEK (30 mM HEPES-KOH, $5\,{\rm mM}$ MgSO$_{\rm 4}$, $1\,{\rm mM}$ DTT, $1\,{\rm mM}$ EGTA and $50\,{\rm mM}$ potassium acetate, pH 7.4) and demembranated with $0.1\%$ (v/v) Igpal. All reagents were purchased from Sigma Aldrich, MO. The membrane-free axonemes were then resuspended in HMDEKP: HMDEK plus $1\%$ (w/v) polyethylene glycol (molecular weight 20 kDa)
with $30\%$ sucrose and stored at $-80\,^{\circ}\mathrm{C}$. After thawing at room temperature, the axonemes were kept on ice and used for up to $4\,{\rm hr}$.

Disintegration assays were performed in flow chambers of depth $100\,\mu {\rm m}$ built from cleaned glass and double-sided sticky tape \citep{alper2012reconstitution}. Thawed axonemes, diluted in HMDEKP, were infused into the chamber and allowed to adhere to the glass surface. Disintegration solution (HMDEKP augmented with $100\,\mu{\rm M\, ATP}$, 0.1\,$\mu$g/ml subtilisin, 10\,units/ml creatine kinase, 6  mM creatine phosphate, 1.3\,mM CaCl$_{\rm 2}$) was then applied for 10 min. After axoneme disintegration, the channel was perfused with reactivation buffer (HMDEKP augmented with 10$\mu$M ATP, 10 units/ml creatine kinase, 6\,mM creatine phosphate, 1.3\,mM CaCl$_{\rm 2}$). The  concentration of free calcium was 0.29\,mM (maxchelator software package v2.2b, {\it http://maxchelator.stanford.edu}). The creatine-based ATP-regeneration system was used to maintain the ATP concentration. 

The disintegrated axonemes were imaged using dark-field microscopy on an inverted Zeiss Axiovert 200 microscope using a 100X iris objective (NA 0.7-1.3) and a dark-field oil condenser (NA 1.4). Movies were recorded at a frame rate of 10 fps using an Andor iXon EM-CCD camera. The filament shape was tracked using custom-built Matlab software. In brief, the filament centerline was obtained by fitting Gaussians to the intensity profile of 250-nm-spaced cross-sections perpendicular to the filament. The starting point was chosen to be on the intact base of the split axoneme. The tangent angle was calculated from the slope of the centerline with respect to the horizontal axis of the image.

\section*{RESULTS}
\section*{Observation of circular arcs in split axonemes}
Axonemes were isolated from the single-celled alga \textit{Chlamydomonas reinhardtii} and demembranated as described in Materials and Methods \citep{alper2012reconstitution}. Treatment with 0.1 $\mu$g/ml subtilisin for 10 min partially disintegrated the axonemes into individual doublets or bundles of doublets, which remained connected at the basal end where the axoneme had been joined to the cell body. The basal end attached to the cover slip. The length of the disintegrated part varied from 5 to 9 $\mu$m.

\begin{Figure}
\begin{center}
\includegraphics [width=8.6cm] {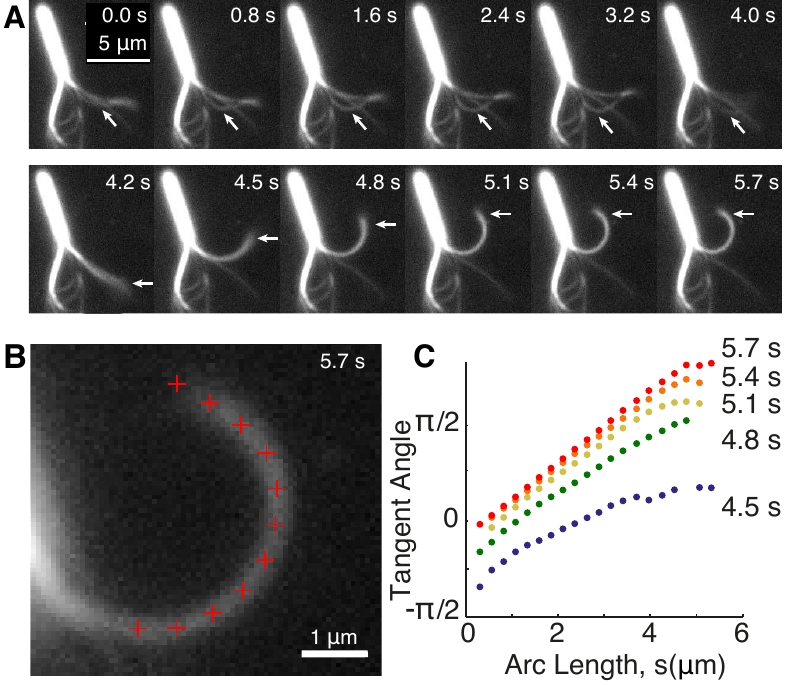}
\captionof{figure}{Sliding of adjacent doublets in a split axoneme (A) One doublet slid along another and then dissociated (0.0 to 4.0\,s). After reassociating (4.2 s), the two doublets remained in close apposition and bent into a circular arc (4.5 s to 5.7 s) before dissociating again (not shown). See also Supplementary Movie 1. (B) The shape was characterized by tracking the centerline of intensity along the filament contour marked by red crosses. The tangent angle was measured between neighboring points. (C) The tangent angles are plotted as a function of arc length, starting from the base. Except at the distal end, the  tangent angle increased linearly with arc length (times 5.1-5.7\,s), indicating that the shape of the doublets is nearly a circular arc. The increasing slope in consecutive frames indicates that the curvature of the arc is increasing over time and approaches a final, quasi-static shape at 5.7\,s.}
\label{fig:bendingmont}
\end{center}
\end{Figure}

At low concentrations of ATP ($10\,\mu {\rm M}$), pairs of filaments partially associated, and propagated small bending waves towards the basal end as one filament slid along the other (Figure \ref{fig:bendingmont}A, first row, arrows). Similar sliding events were reported by \citep{aoyama_cyclical_2005}. We observed a novel property that was not reported previously: sometimes, the two filaments re-associated with each other along their entire length and bent into a circular arc (Figure \ref{fig:bendingmont}A, second row). The system then became unstable and the filaments separated again. This process could repeat itself several times (Supplementary Movie 1 and 2).
{ Note that all the interactions of the filaments including association, bending and dissociation occur in the plane of the image in a chamber that is about 100\,$\mu$m thick, excluding the possibility that the observed arcs are the result of helices in the third dimension flattened by the chamber.}

To analyze the bending in detail, we digitized the shape of the pairs of filaments from the images (Figure \ref{fig:bendingmont}B). We then calculated the tangent angle as a function of arc length in successive frames as the filaments became more and more bent (Figure \ref{fig:bendingmont}C). The key finding was that the filament pair approached a steady-state shape in which the tangent angle increased linearly with arc length, except at the very distal end (Figure \ref{fig:bendingmont}C, $5.1\,{\rm s}$ to $5.7\,{\rm s}$). Such a linearly increasing tangent angle implies that the steady-state shape is approximately a circular arc.

{
We believe that the two interacting filaments are two doublet microtubules and not two singlet microtubules or one doublet microtubule interacting with the central pair. First, by measuring the intensity of the filaments in a partially disintegrated axoneme under conditions designed to observe the central pair, we find that the highly curved central pair has a similar intensity to doublet microtubules under the dark-field microscope (Supplementary Figure 1). This implies that the two individual singlet microtubules that comprise the central pair will each have a lower intensity than a doublet. However, the interacting filaments (Figure 3A,B, red) have the same intensities as the non-interacting filaments (Figure 3A,B, blue), implying that the interacting filaments are not two singlet microtubules (which would both be much dimmer than the non-interacting filaments) or a singlet and a doublet (one of the two interacting filaments would be much dimmer than the other). Note that in dark-field microscopy, when two equally intense filaments come together, the intensity of the pair is about four times that of the individual filaments (Figure 3A, B green) because of the light scattering from two thin, close objects is coherent. This implies that, for example, if the two singlet microtubules have equal mass per unit length, then their individual intensities will be only one quarter that of the central pair. Second, the extruded central pair has high curvature 

(Hosokawa, 1987; Johnson et al., 1994; Lechtreck and Witman, 2007, Supplementary Figure 1). Therefore, the interaction is not between a doublet and a central pair because both of the filaments are only weakly curved when not associated with each other. Third, the speed of sliding of one filament with respect to the other is 2\,$\mu$m/s (Supplementary Movie 2), which, at an ATP concentration of only 10\,$\mu$M, is approximately ten times faster than kinesins \citep{howard} but similar to the speed of some dyneins \citep{Kagami1992}. This suggests that the sliding is not due to kinesins that are associated with the central pair \citep{Johnson:1994ub, Mitchell:1999wf}, but rather to dyneins. Taken together, these arguments suggest that the two filaments are doublet microtubules whose sliding is driven by dynein motors. A similar conclusion was reached by \cite{aoyama_cyclical_2005}.
}
\begin{Figure}
\begin{center}
\includegraphics [width=8.6cm] {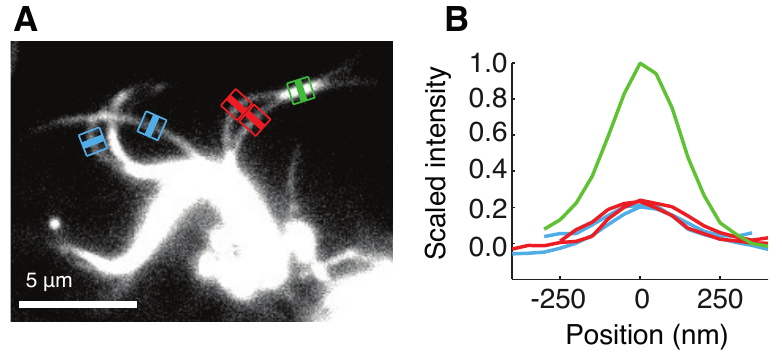}
\captionof{figure}{Intensity analysis of filaments in partially disintegrated axonemes. (A) Frame of a bending event with line-scans of interacting filaments (red), non-interacting filaments (blue) and overlap region of two filaments (green). (B) Corresponding intensity profiles of the line-scans. As shown in A, the interacting- (red) and non-interacting filaments (blue)  have similar intensity. Note that the intensity in the overlap region is approximately 4 times higher than the non-overlap regions. When normalized with respect to the peak intensity of overlapping filaments, the intensities of the individual filaments that did not interact was 0.24$\pm$0.083 (mean$\pm$SD, N=11) compared to 0.29$\pm$0.063 (mean$\pm$SD, N=10) for those filaments that did interact (5 axonemes). There was no statistically significant difference at a 95\% confidence level (t-test).}
\label{fig:intensity}
\end{center}
\end{Figure}

 Assuming that the shape is due to sliding forces between the doublets we can infer the distribution of the active motors from the shape using the moment balance equation derived in the Materials and Methods section:
\begin{align}
\label{moment balance}
F(s)=\kappa C(s)/a_0\quad .
\end{align}
Before applying this equation, however, we have to check that the bent filament pair is indeed in a static mechanical equilibrium in which the frictional forces due to motion through the fluid can be ignored. This assumption is valid in our case, as demonstrated by the following calculation. The characteristic time of the relaxation to the steady state in (Figure \ref{fig:bendingmont}C) is approximately 1 s. The relaxation time of a beam of length ${L}$, stiffness ${\kappa} $ and drag coefficient $\xi_\perp$ is approximately ${ (0.54)^4\xi_\perp L^4}/{\kappa} $ \citep{howard}. The appropriate numerical values are $L= 5\, \mu {\rm m}$, $\xi_\perp= 0.003\, {\rm pN}\cdotp{\rm s}\mu{\rm m}^{-2}$ \citep{riedelkruse_how_2007} and $\kappa=120$ pN$\cdotp \mu{\rm m}^2$. Note that we calculated the bending stiffness of the pair of doublets as follows: the bending stiffness of a single microtubule is 23 pN$\mu$m$^2$ \citep{howard} and the stiffness of one doublet is approximately three times that of a single microtubule \citep{howard}. Assuming $60\, {\rm pN}\cdotp\mu{\rm m}^2$  for each doublet, the rigidity of the pair of doublets is therefore $\kappa=120\,{\rm pN}\cdotp\mu{\rm m}^2$. Using these parameters, we estimate the relaxation time to be $0.001\, {\rm s} \ll 1\, {\rm s} $. Thus, the shape can be considered to be static and the theory applies.   
  
The moment balance equation (Eq. \ref{moment balance}) implies that the motor activity is concentrated at the distal end of the doublets. This follows because a circular shape (i.e. constant curvature) requires a constant total force, which in turn requires the force per unit length $f(s)$ to be zero except at the end where $s=L$, see Figure \ref{fig:modelcomp}AB, blue and green lines. By contrast, if the motors are active all along the doublets (a constant force per unit length $f_{\rm m}(s)=f_{\rm m}$) and there are no crosslinkers opposing shear ($k=0$, as expected given that the doublets can completely separate) then the corresponding total force is $F(s)=(L-s)f_{\rm m}$. This will produce a linearly decreasing curvature $C(s)=(L-s)f_{\rm m}a_0/\kappa$, a spiral, which is clearly inconsistent with the experimental data (compare the red line with the filled circles in Figure \ref{fig:modelcomp}B).

The moment balance equation Eq.\,\ref{moment balance} can also be used to deduce the numerical value of the force generated by the motors to maintain the bent shape of the doublets.  Because the observed curvature of the bent pair in Figure \ref{fig:bendingmont} approaches a constant value $C=0.5 \,\mu {\rm m}^{-1}$, the bending moment at the distal end is $M =\kappa C=60\, {\rm pN}\cdotp\mu{\rm m}$, using the value for $\kappa$ above. If we now assume a spacing between filaments of $a_0=60\,{\rm nm} $ \citep{nicastro_molecular_2006-1}, the shear force generated by the motors  is $F =\kappa C/a_0=1000\, {\rm pN}$.  From Figure \ref{fig:bendingmont}C and Figure \ref{fig:modelcomp}B we infer that all the shear force is produced in the last $0.5-1\,\mu{\rm m}$, over which the curvature drops to zero. This is an unexpectedly high force density; we will return to this point in the Discussion. The force generated by the motors at the distal end must be balanced by the stress in the basal region: $k_{\rm 0} \Delta_{\rm 0}=F_{\rm 0} = 1000\,{\rm pN}$. The implications of this result will also be discussed later.

Why do shear forces accumulate at the distal tip? { One possibility is that the observations are due to a coincidental digestion pattern resulting in the deactivation of all motors away from the distal tip. However, this possibility is not consistent with the non-circular spiral shape observed at early times in the bending (e.g. Figure \ref{fig:bendingmont}AC, 4.5\,s) which implies a broad distribution of motors. In addition, the overlap observed in the buckling phase (i.e. after bending) is often much larger than 1 $\mu$m (see Supplementary Movie 2).} In the following section, we introduce a model that can account for the inferred force distribution. In this model, the activity of motors are regulated either by curvature directly or by the normal forces induced by curvature (Eq. \ref{n-force}).

\begin{Figure}
\begin{center}
\includegraphics [width=8.6cm] {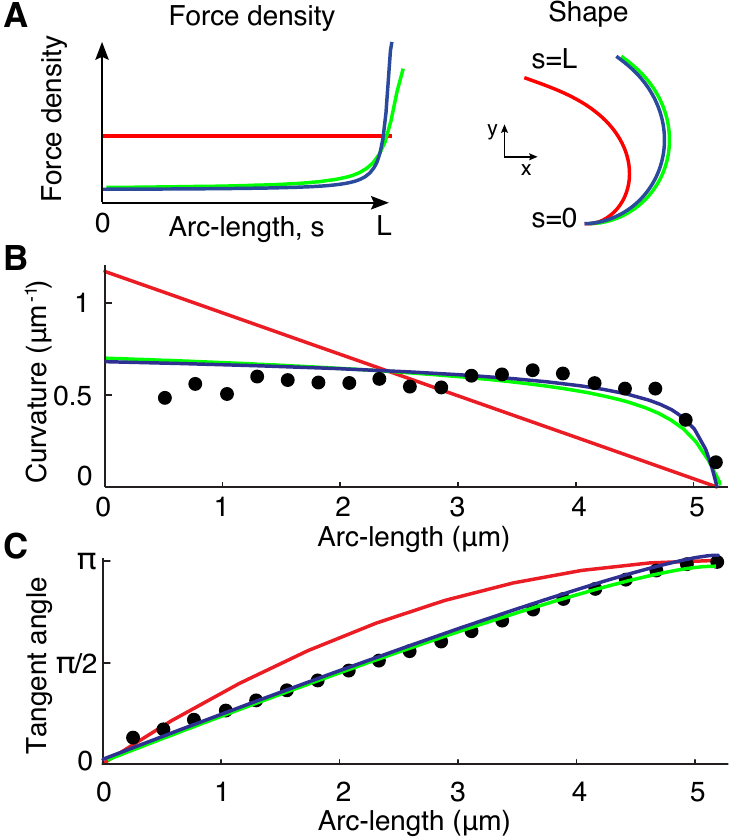}
\captionof{figure}{Comparison of motor-distribution models to the observed filament shape (A) A uniform force distribution (left panel, red curve) produces a spiral shape (right panel, red curve). Tip-concentrated force distributions (left panel, blue and green curves) produce nearly circular arcs (right panel, blue and green curves). (B) and (C)  Curvature and tangent angle data (black dots) from Figure \ref{fig:bendingmont}B compared to the models. Green lines correspond to curvature control, with $\kappa = 120\, {\rm pN}\cdotp\mu{\rm m}^2$, $a_{0}=60\,{\rm nm}$, $\rho= 200\, \mu {\rm m}^{-1}$, $f_+=5$ pN and $C_{\rm c}=0.25\,\mu{\rm m}^{-1}$. Blue lines correspond to normal-force control with  $f_{\rm c}=200\,{\rm  pN}\cdotp\mu{\rm m}^{-1}$. Red lines correspond to the absence of motor  regulation, with a uniform force density of $f_{\rm m}=500\, {\rm pN}\cdotp\mu{\rm m}^{-1}$. The force density was chosen to match the data for the maximum angle at the filament tip.}
\label{fig:modelcomp}
\end{center}
\end{Figure}

\section*{Motor regulation by curvature and normal forces}
We assume that the density of motors $\rho$ is uniform along the doublets.  A single attached motor exerts a force  $f_+>0$, where the sign follows the convention of Figure \ref{fig:schematic} and Appendix B. Motors stochastically attach and detach from the doublet on which they exert force, and $p(s)$ is the probability that a motor at position $s$ is attached. In the absence of cross-linkers (that is, $k=0$), the net force density is  $f(s) = \rho  f_+ p (s)$. For a static configuration, the probability of a motor being attached is given by $p (s) = (1+k_{{\rm off}}(s)/k_{{\rm on}})^{-1}$, with $k_{{\rm on}}$ ($k_{{\rm off}}$) the attachment (detachment) rate. { In this model, the motors are either engaged (and generating force) or not engaged (and not generating force). The disengaged state might correspond to a weak binding state.}

In general, the detachment rate of motors can depend on the forces to which they are subject. { As shown in Eq. \ref{n-force}, the motors are subject to a normal force arising from the filament bending.} { If the motors are sensitive to the normal force and the dependence of the motor detachment rate} follow Bell's law \citep{bell_models_1978}, we have:
\begin{align}
\label{eq:konkofffn}
k_{{\rm off}}(s) = \bar{k}_{{\rm off}} \exp\left[  \frac{f_{\rm n}(s)}{f_{\rm c}}\right]\quad , 
\end{align}
with  $f_c$ the characteristic force density above which normal force significantly increases detachment. Combining force balance  $\delta U /\delta \psi = 0$  with this mechanism gives the equation:
\begin{align}
\label{eq:completefn}
\kappa C'(s)  &= - \frac{\rho a_0 f_+}{1+\frac{\bar{k}_{{\rm off}}}{k_{{\rm on}}}\exp{\left(\frac{C(s)}{ \sqrt{a_0 f_{\rm c}/\kappa}}\right)^2}}\quad .
\end{align}

This equation leads to force concentration at the distal end, as can be appreciated by the following qualitative argument. { Motor sliding forces cause bending. Bending results in a normal force that tends to separate the filament pair. When the normal force exceeds the characteristic normal force density, the motors detach, resulting in a decrease in sliding force.} 
Only near the distal end, at which the curvature decreases to zero (according to the boundary condition), will the curvature fall below the critical curvature and the motors will remain attached. Thus, this motor regulation mechanism results in feedback: as the doublet starts to bend, the higher curvature at the base (Figure \ref{fig:modelcomp}A red curve) causes basal motors to detach, and as the bend develops there will be a wave of detachment that only stops at the distal end.

To compare quantitatively the predictions of this model with the experimental data, we numerically integrated Eq. \ref{eq:completefn} using the boundary condition $C(L)=0$. We used the parameters from the previous sections: motor density $\rho=200\,\mu{\rm m}^{-1}$, and single-motor force $f_+=5\, {\rm pN}$. A characteristic force density $f_{ \rm c}=200\,{\rm pN}\cdotp\mu{\rm m}^{-1}$ was used. In Figure \ref{fig:modelcomp}, we compare the resulting shapes for these parameters (blue lines for normal-force control), with a model where no motor regulation exists (red lines), one with curvature control regulation (green lines, explained below), and the measured data (filled black circles). As plotted in Figure \ref{fig:modelcomp}A, normal-force regulation concentrates the forces at the distal tip. In this region where motors are attached, there is a sharp decrease of the curvature, which is zero at the end of the filament (Figure \ref{fig:modelcomp}B). Figure \ref{fig:modelcomp}C  shows that the normal-force model is in good agreement with the data.

If there is no regulation by normal force, then all motors will have the same probability of being attached, resulting in a constant force per unit length. As already discussed, this results in a linearly decreasing curvature (see Figure \ref{fig:modelcomp}AB, red lines), which is not consistent with the experimental data.  The sliding-force model, where detachment is proportional to the shear force, also leads to a constant force density because, at steady state, the shear force experienced by all the motors is the same, and corresponds to the stall force. Thus, the sliding-force model also has a constant shear force per unit length, and, like the unregulated case, leads to a non-circular shape, inconsistent with the observations. 

In curvature control, the detachment of motors follows a generalization of Bell's law of the form $k_{{\rm off}} = \bar{k}_{{\rm off}} \exp\left[  C/C_{\rm c}\right]$, where $C_{\rm c}$ is the characteristic curvature above which the detachment is significantly enhanced. This curvature dependence results in an equation for the filament's shape analogous to Eq. \ref{eq:completefn}, except with a linear instead of squared dependence in the exponential. Such a model is consistent with the experimental data (see Figure \ref{fig:modelcomp}, green lines). However, while both normal force and curvature control models are consistent with the data, we prefer the normal force model (see Discussion).

\section*{Scaling of curvature with filament length}
{ To determine how curvature depends on filament length, we analyzed five pairs of microtubule doublets that showed the arcing behavior whose lengths ranged from 5 to $9\,\mu$m. A total of 24 arcing events were observed, with up to 8 events for a single pair of filaments (see e.g. Supplementary Movie 2).} 
All bent into nearly circular arcs (Figure \ref{fig:curvlength}, upper images). The average curvature (excluding the last 1 $\mu$m from the distal tip) increased only weakly with the length of the doublets (Figure \ref{fig:curvlength}, filled points). The same parameters used in Figure \ref{fig:modelcomp} were then used to fit the curvature vs. length data for all five doublet pairs, without any additional parameters. Both the curvature control and the normal-force control models were in very good agreement with the data (Figure \ref{fig:curvlength}, green and blue curves respectively). The models predict a weak dependence of the curvature on filament length because it is only the most-distal motors that generate the bending forces. By contrast, if the density of active motors were constant along the doublets, as in the shear-force model, then the average curvature would be proportional to length (integrating Eq. \ref{dgdr}), which is inconsistent with the data. 

\begin{Figure}
\begin{center}
\includegraphics [width=8.6cm] {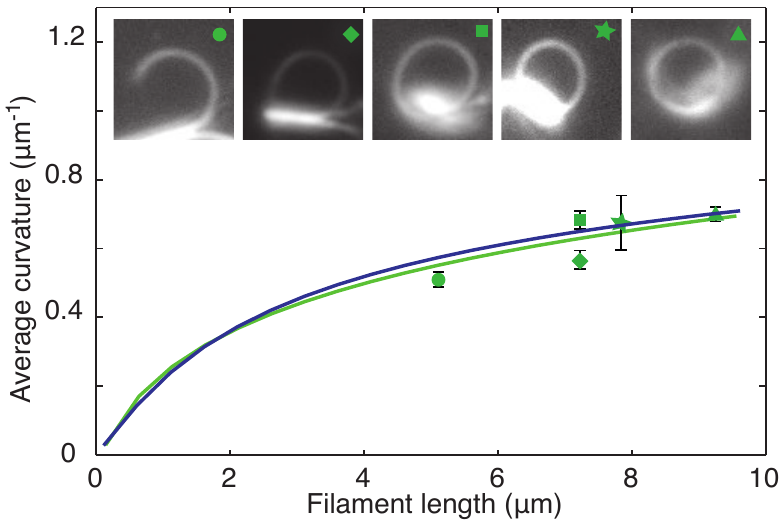}
\captionof{figure}{Dependence of average curvature on filament length. The curvatures of the five tracked split axonemes (insets) show a weak dependence on their length (green symbols). Curvature control (green line) and normal-force control (blue line) both predict the observed weak length dependence, without introducing additional parameters.}
\label{fig:curvlength}
\end{center}
\end{Figure}

\section*{DISCUSSION}
In this work, we partially disintegrated axonemes and found that adjacent doublets underwent cycles of association, dissociation and re-association: while associated, the pairs of doublets bent into circular arcs. There are a number of interesting implications of these observations.

\section*{Nexin crosslinks are not required for static bending}

While dissociated, the doublets are clearly separated by up to a micrometer. Therefore, any permanent links along the doublets, such as nexin, must be absent, presumably digested by the protease. Making the reasonable assumption that the cross links do not reform when the doublets re-associate, our observations imply that lateral crosslinks are not required for static bending and, { indeed, 
our theoretical analysis shows that basal constraints are sufficient for static bending. Earlier work has shown that basal constraints are also sufficient for dynamic bending \citep{riedelkruse_how_2007}. Whether lateral crosslinkers are sufficient for dynamic bending as proposed in some text books \citep{alberts_molecular_2002,phillips2010physical} is still an open question.}  

\section*{The tip-concentrated forces are large}

Our theoretical analysis shows that the observed, nearly circular arcs imply that motor forces are concentrated in the last micrometer (or less) of the distal tip. To bend a pair of doublet microtubules of flexural rigidity $\kappa=120 \, {\rm pN}\cdotp\mu{\rm m}^2$ and separation $a_0=60\,{\rm  nm} $ into an arc of radius, $1/C=2\, \mu{\rm m}$, requires a tip-concentrated shear force of $F =\kappa C/a_0=1000\, {\rm pN}$. 

{     
This force is large considering the number of dyneins per unit length. There are 19 dynein motor domains in each 96 nm structural repeat along a doublet \citep{nicastro_molecular_2006-1}. This corresponds to a density of approximately $200\,\mu{\rm m}^{-1}$. Therefore, if all the dyneins in the last 0.5-1\,$\mu{\rm m}$ were generating force, the average load-force per motor domain would be 5-10\,pN. It is unlikely that the force per motor domain is much less than 5\,pN: the curvature is accurate to within 10\%; the separation cannot be less than 40\,nm based on steric constraints; it is unlikely that the rigidity is much smaller than $\kappa=120 \, {\rm pN}\cdotp\mu{\rm m}^2$, a value that is consistent with both in vivo measurement from sea urchin sperm as well as with in vitro measurements from microtubules \citep{howard}; and the decrease in curvature takes place over a distance that is certainly less than 1\,$\mu{\rm m}$. 

A force of 5\,pN/motor domain presents a puzzle. First, this force is larger than the forces generated per motor domain by the highly processive motors cytoplasmic dynein \citep{gennerich2007force} and kinesin \citep{carter2005mechanics}. Second, it is at the upper limit of estimates from intact axonemes and purified axonemal dyneins (see \cite{Lindemann2003} for references). And third it is close to the thermodynamic limit of 12 pN, which is the maximum possible force if movement through d = 8 nm (the length of the tubulin dimer) and ATP hydrolysis (with the hydrolysis free energy difference $\Delta G_{ATP}=25k_{\rm B}T$, where $k_{\rm B}$ is the Boltzmann constant and T is absolute temperature) are at thermodynamic equilibrium such that $f_+ d=\Delta G_{ATP}$  \citep{howard}. This last calculation suggests that if the step size is 8 nm, then a large fraction of the motor domains are actively generating force: i.e. the duty ratio is high. A low fraction of active motor domains could only mean that axonemal dynein takes ATP-driven steps that are smaller than 8\,nm. Perhaps, each cycle of ATP hydrolysis leads to small rotation of AAA ring and a corresponding small displacement relative to the microtubule. For example, a 10 degree rotation of 13-nm-diameter ring results in a displacement of about 1\,nm and could theoretically take place even against a force as high as 100\,pN. Thus a force of\,5\,pN/motor domain, while not impossible on physical grounds, might suggest that axonemal dyneins have specializations, such as small step sizes, that adapt them to produce high force in the axoneme.}

\section*{Mechanism of motor-regulation}

The concentration of forces at the distal end is intriguing. In the absence of feedback, we would expect sliding motors to build up a force equal to their stall force all along the length of the doublet pair: with a basal constraint, the structure would bend into a spiral, not circular shape. If the motors were simply regulated by the sliding forces, then all motors would be regulated to the same extent and the activity would be constant along the length, again giving rise to a spiral shape. Therefore, it appears that the motors are regulated by curvature directly (curvature-control) or indirectly (normal-force control). 

While curvature control would account for our observations, it is difficult to imagine how dyneins could sense the very small tubulin strains associated with the observed curvatures. A radius of curvature of 2 $\mu{\rm m}$ corresponds to the bending of an 8-nm tubulin dimer through an angle only $0.025^{\circ}$, which is some two orders of magnitude smaller than the curved-to-straight conformation associated with the straightening of a free GDP-bound tubulin subunit needed for its incorporation into the microtubule wall \citep{ravelli2004insight}. Bending a doublet microtubule (of diameter approximately 40 nm) into an arc of radius 2 $\mu{\rm m}$ would stretch a tubulin dimer in the outer wall by 1\%, or less than 0.1\,nm over its 8-nm length. These are small conformational changes and it is difficult to understand how they could be recognized by dynein's microtubule-binding domain.

{
On the other hand, an observed curvature of 0.5 $\mu{\rm m}^{-1}$ produces a normal force of $f_{\rm n}=500\,{\rm pN}\cdotp\mu{\rm m}^{-1}$. Using Eq. \ref{n-force} and our modeling shows that a critical normal force of $200\, {\rm pN}\cdotp\mu{\rm m}^{-1}$, or 1\,pN per motor domain, is sufficient to account for the observed shape. This value of the critical normal force from our static measurements is similar to that estimated by \cite{Lindemann2003} to account for normal force control in the dynamic, beating axoneme. Such a critical normal force would have to be associated with a protein conformational change of about 4\,nm (${\rm =\textit{k$_{\rm B}$T}/1\,pN}$), or less if not all the dynein heads are attached. Because this distance is much smaller than the 30-nm length of a dynein molecule, the normal-force model is more plausible on structural grounds.}

This study has involved ``static", slowly moving axonemes, rather than the dynamic, rapidly moving axonemes in the intact flagellum. Nevertheless, it is tempting to speculate that the normal-force (or perhaps curvature) regulation seen in the static case also applies to the dynamic case. Whether it does or not will require further experiments and calculations. Interestingly, in earlier work we accounted for the dynamic beat of bull sperm axonemes using a sliding-force regulation model \citep{riedelkruse_how_2007}, which does not accord with the results reported here for \textit{Chlamydomonas}. Any possible difference in dynein regulation between these two different structures may be attributed to their very different swimming strokes: \textit{Chlamydomonas} uses a breast-stroke beat which pulls the cell body from the front, while sperm have a snake-like beat that pushes the sperm head from behind. The large curvatures in the breast-stroke may make this a preferred regulatory mechanism, especially considering that the normal force scales with the square of curvature.

\section*{Incompressibility of the doublets}

The last issue to discuss is the incompressibility assumption. The total force acting at the base is 1000 pN. Such a force, acting on a doublet microtubule of length $5\,  \mu{\rm m}$, cross-sectional area $400\, {\rm nm}^2$ \citep{howard} and Young's modulus 2 GPa, would cause a relative compression of about 0.1\% or $5\, {\rm nm}$. This is very much smaller than the sliding displacement of 150 nm for a radius of curvature of $2\,  \mu{\rm m}$ and a doublet separation of $60\,{\rm nm}$ (Eq. \ref{slidingd}). In other words, the shear force is small enough that the pair of doublets can still be considered as inextensible filaments undergoing shear displacements.

\section*{Appendix A: Variations}
The total variation of the internal energy from Eq. \ref{FreeEnergy} has three contributions $\delta U = \delta U_\psi+ \delta U_a+ \delta U_0$ respectively corresponding to variations of the angle $\psi(s)$, the spacing $a(s)$, and the basal sliding $\Delta_{\rm 0}$. For the first term $\delta U_\psi$ the angle is varied:
\begin{align}
\delta  U_\psi = \int_0^L\left\{\kappa\dot{\psi}(s)\delta\dot{\psi}(s) -f(s)\int_0^s[a(s')\delta\dot{\psi}(s')]\d s'\right\}\d s
\end{align}
where we have used the definition of shear force density $f(s)=f_{\rm m}(s)- k\Delta(s)$ and the geometric relation in Eq. \ref{slidingd}. We integrate the first term by parts to remove the arclength derivative from the angle variation, and also the second to get the variation out of the integral. This results in:
\begin{align}
\delta  U_\psi& = \left[\kappa\dot{\psi}(s)\delta\psi(s)\right]_0^L
-\left[ \int_0^{s}f(s')\d s' \int_0^{s}a(s')\delta\dot{\psi}(s')\d s' \right]_0^L
\nonumber\\
&+\int_0^L\left\{-\kappa\ddot{\psi}(s)\delta\psi(s)
+a(s)\delta\dot{\psi}(s)\int_0^sf(s')\d s'  \right\}\d s\nonumber\\
&=\left[\kappa\dot{\psi}(s)\delta\psi(s)\right]_0^L\nonumber\\
&+\int_0^L\left\{-\kappa\ddot{\psi}(s)\delta\psi(s)-a(s)F(s)\delta\dot{\psi}(s) \right\}\d s
\end{align}
%I corrected a sign error (please check)
Where in the last equality we have introduced the second boundary term in the bulk integral, and used the definition of integrated shear force $F(s)=\int_s^Lf(s')\d s'$. Integrating again by parts to remove the derivative leads to:
\begin{align}
\delta  U_\psi& =\left[(\kappa\dot{\psi}(s) -a(s) F(s))\delta\psi(s)\right]_0^L\nonumber\\  
&+\int_0^L\left\{-\kappa\ddot{\psi}(s)
-a(s)f(s)+\dot{a}(s)F(s) \right\}\delta\psi(s)\d s
\end{align}
%I corrected a sign error (please check)
Where the boundary terms represent the boundary torques. In the case of an isolated system with no external torques applied we thus have as boundary conditions
\begin{align}
\dot{\psi}(L)&=0\nonumber\\
\kappa\dot{\psi}(0) -a(0) F(0)&=0
\end{align}

The second term $\delta U_2$ is obtained by varying the spacing $a(s)$:
\begin{align}
\delta  U_a = \int_0^L\left\{-f(s)\int_0^s\dot{\psi}(s')\delta a(s')\d s'+f_{\rm n}(s)\delta a(s)\right\}\d s
\end{align}
where $f_{\rm n}(s)=k_{\rm n}(a(s)-a_0)$ has been introduced. Integrating by parts results in:
\begin{align}
\delta  U_a = \left[-\int_0^{s}f(s')\d s'\int_0^{s}\dot{\psi}(s')\delta a(s')\d s'\right]_0^L\nonumber\\
+\int_0^L\left\{\dot{\psi}(s)\int_0^s f(s')\d s' +f_{\rm n}(s)\right\}\delta a(s)\d s
\end{align}
Which after introducing the boundary term in the bulk integral and using the definition of $F(s)$, results in:
\begin{align}
\delta  U_a =\int_0^L\left\{ f_{\rm n}(s) - \dot{\psi}(s)F(s) \right\}\delta a(s)\d s 
\end{align}
Notice that this variational term has no boundary contributions.

Finally, the third term $\delta U_0$ is obtained by doing variations with respect to the basal sliding $\Delta _{\rm 0}$:
\begin{align}
\delta  U_0 =-\int_0^L f(s)  \delta \Delta_0\d s +F_{\rm 0}\delta\Delta_{\rm 0}
\end{align}
Where we introduced $F_{\rm 0}=k_{\rm 0}\Delta_{\rm 0}$. Since $\Delta_{\rm 0}$ does not depend on the arc-length, we get
\begin{align}
\delta  U_0 =(F_{\rm 0}-F(0))\delta\Delta_{\rm 0}
\end{align}
Where we have used the definition of $F(s)$.

\section*{Appendix B: Sign convention}
The sign convention used throughout this paper is as follows. Positive motor force densities $f_{\rm m}(s)$ are opposed by positive shearing forces $k\Delta(s)$ to generate a net positive shear force, that is $f(s)=f_{\rm m}(s)-k\Delta(s)>0$. With this convention in Figure 1 the shearing is positive, as is the angle and the curvature. The integrated force $F(s)=\int_s^Lf(s')\d s'$ is also positive, and at the base it is balanced by the positive basal force $F_0=k_0\Delta_0$ such that $F(s=0)=k_0\Delta_0$. This means that also the basal sliding is positive. Finally, as at the distal end the curvature vanishes $C(s=L)=0$, we have by Eq. \ref{eq:completefn} that the curvature is a decreasing quantity over arc-length (it's derivative is negative).

\bibliographystyle{aipauth4-1}
\bibliography{SplitAxoneme}

%merlin.mbs aipauth4-1.bst 2010-07-25 4.21a (PWD, AO, DPC) hacked
%Control: author (9) reversed initials
\begin{thebibliography}{26}%
\makeatletter
\providecommand \@ifxundefined [1]{%
 \@ifx{#1\undefined}
}%
\providecommand \@ifnum [1]{%
 \ifnum #1\expandafter \@firstoftwo
 \else \expandafter \@secondoftwo
 \fi
}%
\providecommand \@ifx [1]{%
 \ifx #1\expandafter \@firstoftwo
 \else \expandafter \@secondoftwo
 \fi
}%
\providecommand \natexlab [1]{#1}%
\providecommand \enquote  [1]{``#1''}%
\providecommand \bibnamefont  [1]{#1}%
\providecommand \bibfnamefont [1]{#1}%
\providecommand \citenamefont [1]{#1}%
\providecommand \href@noop [0]{\@secondoftwo}%
\providecommand \href [0]{\begingroup \@sanitize@url \@href}%
\providecommand \@href[1]{\@@startlink{#1}\@@href}%
\providecommand \@@href[1]{\endgroup#1\@@endlink}%
\providecommand \@sanitize@url [0]{\catcode `\\12\catcode `\$12\catcode
  `\&12\catcode `\#12\catcode `\^12\catcode `\_12\catcode `\%12\relax}%
\providecommand \@@startlink[1]{}%
\providecommand \@@endlink[0]{}%
\providecommand \url  [0]{\begingroup\@sanitize@url \@url }%
\providecommand \@url [1]{\endgroup\@href {#1}{\urlprefix }}%
\providecommand \urlprefix  [0]{URL }%
\providecommand \Eprint [0]{\href }%
\providecommand \doibase [0]{http://dx.doi.org/}%
\providecommand \selectlanguage [0]{\@gobble}%
\providecommand \bibinfo  [0]{\@secondoftwo}%
\providecommand \bibfield  [0]{\@secondoftwo}%
\providecommand \translation [1]{[#1]}%
\providecommand \BibitemOpen [0]{}%
\providecommand \bibitemStop [0]{}%
\providecommand \bibitemNoStop [0]{.\EOS\space}%
\providecommand \EOS [0]{\spacefactor3000\relax}%
\providecommand \BibitemShut  [1]{\csname bibitem#1\endcsname}%
\let\auto@bib@innerbib\@empty
%</preamble>
\bibitem [{\citenamefont {{Alberts et al}}(2002)}]{alberts_molecular_2002}%
  \BibitemOpen
  \bibfield  {author} {\bibinfo {author} {\bibnamefont {{Alberts et al}},
  \bibfnamefont {B.}},\ }\href@noop {} {\bibfield  {journal} {\bibinfo
  {journal} {Garland Science, New York}\ } (\bibinfo {year}
  {2002})}\BibitemShut {NoStop}%
\bibitem [{\citenamefont {Alper}\ \emph {et~al.}(2013)\citenamefont {Alper},
  \citenamefont {Geyer}, \citenamefont {Mukundan}, \citenamefont {Howard} \emph
  {et~al.}}]{alper2012reconstitution}%
  \BibitemOpen
  \bibfield  {author} {\bibinfo {author} {\bibnamefont {Alper}, \bibfnamefont
  {J.}}, \bibinfo {author} {\bibnamefont {Geyer}, \bibfnamefont {V.}}, \bibinfo
  {author} {\bibnamefont {Mukundan}, \bibfnamefont {V.}}, \bibinfo {author}
  {\bibnamefont {Howard}, \bibfnamefont {J.}},  \emph {et~al.},\ }\href@noop {}
  {\bibfield  {journal} {\bibinfo  {journal} {Methods in enzymology}\ }\textbf
  {\bibinfo {volume} {524}},\ \bibinfo {pages} {343} (\bibinfo {year}
  {2013})}\BibitemShut {NoStop}%
\bibitem [{\citenamefont {Aoyama}\ and\ \citenamefont
  {Kamiya}(2005)}]{aoyama_cyclical_2005}%
  \BibitemOpen
  \bibfield  {author} {\bibinfo {author} {\bibnamefont {Aoyama}, \bibfnamefont
  {S.}}\ and\ \bibinfo {author} {\bibnamefont {Kamiya}, \bibfnamefont {R.}},\
  }\href {\doibase 10.1529/biophysj.105.067876} {\bibfield  {journal} {\bibinfo
   {journal} {Biophysical Journal}\ }\textbf {\bibinfo {volume} {89}},\
  \bibinfo {pages} {3261} (\bibinfo {year} {2005})}\BibitemShut {NoStop}%
\bibitem [{\citenamefont {Bell}(1978)}]{bell_models_1978}%
  \BibitemOpen
  \bibfield  {author} {\bibinfo {author} {\bibnamefont {Bell}, \bibfnamefont
  {G.}},\ }\href {\doibase 10.1126/science.2928796} {\bibfield  {journal}
  {\bibinfo  {journal} {Science}\ }\textbf {\bibinfo {volume} {243}},\ \bibinfo
  {pages} {618 } (\bibinfo {year} {1978})}\BibitemShut {NoStop}%
\bibitem [{\citenamefont {Brokaw}(1989)}]{brokaw_direct_1989}%
  \BibitemOpen
  \bibfield  {author} {\bibinfo {author} {\bibnamefont {Brokaw}, \bibfnamefont
  {C.}},\ }\href {\doibase 10.1126/science.2928796} {\bibfield  {journal}
  {\bibinfo  {journal} {Science}\ }\textbf {\bibinfo {volume} {243}},\ \bibinfo
  {pages} {1593 } (\bibinfo {year} {1989})}\BibitemShut {NoStop}%
\bibitem [{\citenamefont {Brokaw}(1971)}]{brokaw_bend_1971}%
  \BibitemOpen
  \bibfield  {author} {\bibinfo {author} {\bibnamefont {Brokaw}, \bibfnamefont
  {C.~J.}},\ }\href {http://jeb.biologists.org/content/55/2/289} {\bibfield
  {journal} {\bibinfo  {journal} {Journal of Experimental Biology}\ }\textbf
  {\bibinfo {volume} {55}},\ \bibinfo {pages} {289} (\bibinfo {year}
  {1971})}\BibitemShut {NoStop}%
\bibitem [{\citenamefont {Brokaw}(1975)}]{brokaw_molecular_1975}%
  \BibitemOpen
  \bibfield  {author} {\bibinfo {author} {\bibnamefont {Brokaw}, \bibfnamefont
  {C.~J.}},\ }\href {http://www.pnas.org/content/72/8/3102} {\bibfield
  {journal} {\bibinfo  {journal} {Proceedings of the National Academy of
  Sciences}\ }\textbf {\bibinfo {volume} {72}},\ \bibinfo {pages} {3102}
  (\bibinfo {year} {1975})}\BibitemShut {NoStop}%
\bibitem [{\citenamefont {Brokaw}(2009)}]{brokaw_thinking_2009}%
  \BibitemOpen
  \bibfield  {author} {\bibinfo {author} {\bibnamefont {Brokaw}, \bibfnamefont
  {C.~J.}},\ }\href {\doibase 10.1002/cm.20313} {\bibfield  {journal} {\bibinfo
   {journal} {Cell Motility and the Cytoskeleton}\ }\textbf {\bibinfo {volume}
  {66}},\ \bibinfo {pages} {425} (\bibinfo {year} {2009})}\BibitemShut
  {NoStop}%
\bibitem [{\citenamefont {Camalet}\ and\ \citenamefont
  {J\"{u}licher}(2000)}]{camalet_generic_2000}%
  \BibitemOpen
  \bibfield  {author} {\bibinfo {author} {\bibnamefont {Camalet}, \bibfnamefont
  {S.}}\ and\ \bibinfo {author} {\bibnamefont {J\"{u}licher}, \bibfnamefont
  {F.}},\ }\href {\doibase 10.1088/1367-2630/2/1/324} {\bibfield  {journal}
  {\bibinfo  {journal} {New Journal of Physics}\ }\textbf {\bibinfo {volume}
  {2}},\ \bibinfo {pages} {24} (\bibinfo {year} {2000})}\BibitemShut {NoStop}%
\bibitem [{\citenamefont {Carter}\ and\ \citenamefont
  {Cross}(2005)}]{carter2005mechanics}%
  \BibitemOpen
  \bibfield  {author} {\bibinfo {author} {\bibnamefont {Carter}, \bibfnamefont
  {N.~J.}}\ and\ \bibinfo {author} {\bibnamefont {Cross}, \bibfnamefont {R.}},\
  }\href@noop {} {\bibfield  {journal} {\bibinfo  {journal} {Nature}\ }\textbf
  {\bibinfo {volume} {435}},\ \bibinfo {pages} {308} (\bibinfo {year}
  {2005})}\BibitemShut {NoStop}%
\bibitem [{\citenamefont {Gennerich}\ \emph {et~al.}(2007)\citenamefont
  {Gennerich}, \citenamefont {Carter}, \citenamefont {Reck-Peterson},\ and\
  \citenamefont {Vale}}]{gennerich2007force}%
  \BibitemOpen
  \bibfield  {author} {\bibinfo {author} {\bibnamefont {Gennerich},
  \bibfnamefont {A.}}, \bibinfo {author} {\bibnamefont {Carter}, \bibfnamefont
  {A.~P.}}, \bibinfo {author} {\bibnamefont {Reck-Peterson}, \bibfnamefont
  {S.~L.}}, \ and\ \bibinfo {author} {\bibnamefont {Vale}, \bibfnamefont
  {R.~D.}},\ }\href@noop {} {\bibfield  {journal} {\bibinfo  {journal} {Cell}\
  }\textbf {\bibinfo {volume} {131}},\ \bibinfo {pages} {952} (\bibinfo {year}
  {2007})}\BibitemShut {NoStop}%
\bibitem [{\citenamefont {Howard}(2001)}]{howard}%
  \BibitemOpen
  \bibfield  {author} {\bibinfo {author} {\bibnamefont {Howard}, \bibfnamefont
  {J.}},\ }\href@noop {} {\bibfield  {journal} {\bibinfo  {journal} {Sinauer
  Associates, Sunderland, MA}\ } (\bibinfo {year} {2001})}\BibitemShut
  {NoStop}%
\bibitem [{\citenamefont {Johnson}, \citenamefont {Haas},\ and\ \citenamefont
  {Rosenbaum}(1994)}]{Johnson:1994ub}%
  \BibitemOpen
  \bibfield  {author} {\bibinfo {author} {\bibnamefont {Johnson}, \bibfnamefont
  {K.~A.}}, \bibinfo {author} {\bibnamefont {Haas}, \bibfnamefont {M.~A.}}, \
  and\ \bibinfo {author} {\bibnamefont {Rosenbaum}, \bibfnamefont {J.~L.}},\
  }\href@noop {} {\  (\bibinfo {year} {1994})}\BibitemShut {NoStop}%
\bibitem [{\citenamefont {J\"{u}licher}\ and\ \citenamefont
  {Prost}(1997)}]{julicher_spontaneous_1997}%
  \BibitemOpen
  \bibfield  {author} {\bibinfo {author} {\bibnamefont {J\"{u}licher},
  \bibfnamefont {F.}}\ and\ \bibinfo {author} {\bibnamefont {Prost},
  \bibfnamefont {J.}},\ }\href {\doibase 10.1103/PhysRevLett.78.4510}
  {\bibfield  {journal} {\bibinfo  {journal} {Physical Review Letters}\
  }\textbf {\bibinfo {volume} {78}},\ \bibinfo {pages} {4510} (\bibinfo {year}
  {1997})}\BibitemShut {NoStop}%
\bibitem [{\citenamefont {Kagami}\ and\ \citenamefont
  {Kamiya}(1992)}]{Kagami1992}%
  \BibitemOpen
  \bibfield  {author} {\bibinfo {author} {\bibnamefont {Kagami}, \bibfnamefont
  {O.}}\ and\ \bibinfo {author} {\bibnamefont {Kamiya}, \bibfnamefont {R.}},\
  }\href@noop {} {\bibfield  {journal} {\bibinfo  {journal} {Journal of Cell
  Science}\ }\textbf {\bibinfo {volume} {103}},\ \bibinfo {pages} {653}
  (\bibinfo {year} {1992})}\BibitemShut {NoStop}%
\bibitem [{\citenamefont {Lindemann}(1994)}]{lindemann_model_1994}%
  \BibitemOpen
  \bibfield  {author} {\bibinfo {author} {\bibnamefont {Lindemann},
  \bibfnamefont {C.~B.}},\ }\href {\doibase 10.1002/cm.970290206} {\bibfield
  {journal} {\bibinfo  {journal} {Cell Motility and the Cytoskeleton}\ }\textbf
  {\bibinfo {volume} {29}},\ \bibinfo {pages} {141} (\bibinfo {year}
  {1994})}\BibitemShut {NoStop}%
\bibitem [{\citenamefont {Lindemann}(2003)}]{Lindemann2003}%
  \BibitemOpen
  \bibfield  {author} {\bibinfo {author} {\bibnamefont {Lindemann},
  \bibfnamefont {C.~B.}},\ }\href@noop {} {\bibfield  {journal} {\bibinfo
  {journal} {Biophysical Journal}\ }\textbf {\bibinfo {volume} {84}},\ \bibinfo
  {pages} {4115} (\bibinfo {year} {2003})}\BibitemShut {NoStop}%
\bibitem [{\citenamefont {Mitchell}\ and\ \citenamefont
  {Sale}(1999)}]{Mitchell:1999wf}%
  \BibitemOpen
  \bibfield  {author} {\bibinfo {author} {\bibnamefont {Mitchell},
  \bibfnamefont {D.~R.}}\ and\ \bibinfo {author} {\bibnamefont {Sale},
  \bibfnamefont {W.~S.}},\ }\href@noop {} {\bibfield  {journal} {\bibinfo
  {journal} {Journal of Cell Biology}\ }\textbf {\bibinfo {volume} {144}},\
  \bibinfo {pages} {293} (\bibinfo {year} {1999})}\BibitemShut {NoStop}%
\bibitem [{\citenamefont {Nicastro}\ \emph {et~al.}(2006)\citenamefont
  {Nicastro}, \citenamefont {Schwartz}, \citenamefont {Pierson}, \citenamefont
  {Gaudette}, \citenamefont {Porter},\ and\ \citenamefont
  {{McIntosh}}}]{nicastro_molecular_2006-1}%
  \BibitemOpen
  \bibfield  {author} {\bibinfo {author} {\bibnamefont {Nicastro},
  \bibfnamefont {D.}}, \bibinfo {author} {\bibnamefont {Schwartz},
  \bibfnamefont {C.}}, \bibinfo {author} {\bibnamefont {Pierson}, \bibfnamefont
  {J.}}, \bibinfo {author} {\bibnamefont {Gaudette}, \bibfnamefont {R.}},
  \bibinfo {author} {\bibnamefont {Porter}, \bibfnamefont {M.~E.}}, \ and\
  \bibinfo {author} {\bibnamefont {{McIntosh}}, \bibfnamefont {J.~R.}},\
  }\href@noop {} {\bibfield  {journal} {\bibinfo  {journal} {Science}\ }\textbf
  {\bibinfo {volume} {313}},\ \bibinfo {pages} {944} (\bibinfo {year}
  {2006})}\BibitemShut {NoStop}%
\bibitem [{\citenamefont {Okuno}(1980)}]{Okuno1980}%
  \BibitemOpen
  \bibfield  {author} {\bibinfo {author} {\bibnamefont {Okuno}, \bibfnamefont
  {M.}},\ }\href@noop {} {\bibfield  {journal} {\bibinfo  {journal} {Journal of
  Cell Biology}\ }\textbf {\bibinfo {volume} {85}},\ \bibinfo {pages} {712}
  (\bibinfo {year} {1980})}\BibitemShut {NoStop}%
\bibitem [{\citenamefont {Pazour}\ \emph {et~al.}(2005)\citenamefont {Pazour},
  \citenamefont {Agrin}, \citenamefont {Leszyk},\ and\ \citenamefont
  {Witman}}]{pazour_proteomic_2005}%
  \BibitemOpen
  \bibfield  {author} {\bibinfo {author} {\bibnamefont {Pazour}, \bibfnamefont
  {G.~J.}}, \bibinfo {author} {\bibnamefont {Agrin}, \bibfnamefont {N.}},
  \bibinfo {author} {\bibnamefont {Leszyk}, \bibfnamefont {J.}}, \ and\
  \bibinfo {author} {\bibnamefont {Witman}, \bibfnamefont {G.~B.}},\ }\href
  {\doibase 10.1083/jcb.200504008} {\bibfield  {journal} {\bibinfo  {journal}
  {Science}\ }\textbf {\bibinfo {volume} {170}},\ \bibinfo {pages} {103 }
  (\bibinfo {year} {2005})}\BibitemShut {NoStop}%
\bibitem [{\citenamefont {Phillips}\ \emph {et~al.}(2010)\citenamefont
  {Phillips}, \citenamefont {Kondev}, \citenamefont {Theriot}, \citenamefont
  {Garcia}, \citenamefont {Chasan} \emph {et~al.}}]{phillips2010physical}%
  \BibitemOpen
  \bibfield  {author} {\bibinfo {author} {\bibnamefont {Phillips},
  \bibfnamefont {R.}}, \bibinfo {author} {\bibnamefont {Kondev}, \bibfnamefont
  {J.}}, \bibinfo {author} {\bibnamefont {Theriot}, \bibfnamefont {J.}},
  \bibinfo {author} {\bibnamefont {Garcia}, \bibfnamefont {H.}}, \bibinfo
  {author} {\bibnamefont {Chasan}, \bibfnamefont {B.}},  \emph {et~al.},\
  }\href@noop {} {\bibfield  {journal} {\bibinfo  {journal} {American Journal
  of Physics}\ }\textbf {\bibinfo {volume} {78}},\ \bibinfo {pages} {1230}
  (\bibinfo {year} {2010})}\BibitemShut {NoStop}%
\bibitem [{\citenamefont {Ravelli}\ \emph {et~al.}(2004)\citenamefont
  {Ravelli}, \citenamefont {Gigant}, \citenamefont {Curmi}, \citenamefont
  {Jourdain}, \citenamefont {Lachkar}, \citenamefont {Sobel},\ and\
  \citenamefont {Knossow}}]{ravelli2004insight}%
  \BibitemOpen
  \bibfield  {author} {\bibinfo {author} {\bibnamefont {Ravelli}, \bibfnamefont
  {R.~B.}}, \bibinfo {author} {\bibnamefont {Gigant}, \bibfnamefont {B.}},
  \bibinfo {author} {\bibnamefont {Curmi}, \bibfnamefont {P.~A.}}, \bibinfo
  {author} {\bibnamefont {Jourdain}, \bibfnamefont {I.}}, \bibinfo {author}
  {\bibnamefont {Lachkar}, \bibfnamefont {S.}}, \bibinfo {author} {\bibnamefont
  {Sobel}, \bibfnamefont {A.}}, \ and\ \bibinfo {author} {\bibnamefont
  {Knossow}, \bibfnamefont {M.}},\ }\href@noop {} {\bibfield  {journal}
  {\bibinfo  {journal} {Nature}\ }\textbf {\bibinfo {volume} {428}},\ \bibinfo
  {pages} {198} (\bibinfo {year} {2004})}\BibitemShut {NoStop}%
\bibitem [{\citenamefont {{Riedel-Kruse}}\ \emph {et~al.}(2007)\citenamefont
  {{Riedel-Kruse}}, \citenamefont {Hilfinger}, \citenamefont {Howard},\ and\
  \citenamefont {J\"{u}licher}}]{riedelkruse_how_2007}%
  \BibitemOpen
  \bibfield  {author} {\bibinfo {author} {\bibnamefont {{Riedel-Kruse}},
  \bibfnamefont {I.~H.}}, \bibinfo {author} {\bibnamefont {Hilfinger},
  \bibfnamefont {A.}}, \bibinfo {author} {\bibnamefont {Howard}, \bibfnamefont
  {J.}}, \ and\ \bibinfo {author} {\bibnamefont {J\"{u}licher}, \bibfnamefont
  {F.}},\ }\href {\doibase 10.2976/1.2773861} {\bibfield  {journal} {\bibinfo
  {journal} {{HFSP} Journal}\ }\textbf {\bibinfo {volume} {1}},\ \bibinfo
  {pages} {192} (\bibinfo {year} {2007})}\BibitemShut {NoStop}%
\bibitem [{\citenamefont {Summers}\ and\ \citenamefont
  {Gibbons}(1971)}]{summers_adenosine_1971}%
  \BibitemOpen
  \bibfield  {author} {\bibinfo {author} {\bibnamefont {Summers}, \bibfnamefont
  {K.~E.}}\ and\ \bibinfo {author} {\bibnamefont {Gibbons}, \bibfnamefont
  {I.~R.}},\ }\href {http://www.pnas.org/content/68/12/3092} {\bibfield
  {journal} {\bibinfo  {journal} {Proceedings of the National Academy of
  Sciences}\ }\textbf {\bibinfo {volume} {68}},\ \bibinfo {pages} {3092}
  (\bibinfo {year} {1971})}\BibitemShut {NoStop}%
\bibitem [{\citenamefont {Witman}, \citenamefont {Plummer},\ and\ \citenamefont
  {Sander}(1978)}]{witman_chlamydomonas_1978}%
  \BibitemOpen
  \bibfield  {author} {\bibinfo {author} {\bibnamefont {Witman}, \bibfnamefont
  {G.~B.}}, \bibinfo {author} {\bibnamefont {Plummer}, \bibfnamefont {J.}}, \
  and\ \bibinfo {author} {\bibnamefont {Sander}, \bibfnamefont {G.}},\
  }\href@noop {} {\bibfield  {journal} {\bibinfo  {journal} {The Journal of
  Cell Biology}\ }\textbf {\bibinfo {volume} {76}},\ \bibinfo {pages} {729}
  (\bibinfo {year} {1978})},\ \bibinfo {note} {{PMID:} 632325 {PMCID:}
  2110011}\BibitemShut {NoStop}%
\end{thebibliography}%

\end{document}